\documentclass[aps,reprint,superscriptaddress]{revtex4-1}
\usepackage{amsmath}
\usepackage{amssymb}
\usepackage{graphicx}

\usepackage{hyperref}
\usepackage{bm}
\usepackage{color}

\newcommand{\mysize}{3.4}

\begin{document}


\title{Generation of polarized positron beams via collisions of ultrarelativistic electron beams}

\affiliation{Beijing National Laboratory for Condensed Matter Physics, Institute of Physics, CAS, Beijing 100190, China}
\affiliation{Department of Physics and Beijing Key Laboratory of Opto-electronic Functional Materials and Micro–nano Devices, Renmin University of China, Beijing 100872, China}
\affiliation{School of Physical Sciences, University of Chinese Academy of Sciences, Beijing 100049, China}
\affiliation{Songshan Lake Materials Laboratory, Dongguan, Guangdong 523808, China}
\affiliation{CAS Center for Excellence in Ultra-intense Laser Science, Shanghai 201800, China}

\author{Huai-Hang Song}
\affiliation{Beijing National Laboratory for Condensed Matter Physics, Institute of Physics, CAS, Beijing 100190, China}
\affiliation{School of Physical Sciences, University of Chinese Academy of Sciences, Beijing 100049, China}
\author{Wei-Min Wang}\email{weiminwang1@ruc.edu.cn}
\affiliation{Department of Physics and Beijing Key Laboratory of Opto-electronic Functional Materials and Micro–nano Devices, Renmin University of China, Beijing 100872, China}
\affiliation{Beijing National Laboratory for Condensed Matter Physics, Institute of Physics, CAS, Beijing 100190, China}
%
%
\author{Yu-Tong Li}
\email{ytli@iphy.ac.cn}
\affiliation{Beijing National Laboratory for Condensed Matter Physics,
Institute of Physics, CAS, Beijing 100190, China}
\affiliation{School of Physical Sciences, University of Chinese Academy of Sciences, Beijing 100049, China}
\affiliation{Songshan Lake Materials Laboratory, Dongguan, Guangdong 523808, China}
\affiliation{CAS Center for Excellence in Ultra-intense Laser Science, Shanghai 201800, China}

\date{\today}

\begin{abstract}

A novel scheme is proposed for generating a polarized positron beam via multiphoton Breit-Wheeler process during the collision of a 10 GeV, pC seeding electron beam with the other 1 GeV, nC driving electron beam. The driving beam provides the strong self-generated field, and a suitable transverse deviation distance between two beams enables the field experienced by the seeding beam to be unipolar, which is crucial for realizing the positron polarization. We employ the  particle simulation with a Monte-Carlo method to calculate the spin- and polarization-resolved photon emission and electron-positron pair production in the local constant field approximation. Our simulation results show that a highly polarized positron beam with polarization above $40\%$ can be generated in several femtoseconds, which is robust with respect to parameters of two electron beams. Based on an analysis of the influence of $\gamma$-photon polarization on the polarized pair production, we find that a polarized seeding beam of the proper initial polarization can further improve the positron polarization to $60\%$.

\end{abstract}


\maketitle


\section{Introduction}

Polarized positron beams are indispensable tools in many areas of science and technology. In surface physics, low-energy polarized positron beams can be utilized to probe surface magnetism \cite{Gidley1982prl}, current-induced spin polarization \cite{Kawasuso2013jmmm}, and other spin phenomena at surfaces \cite{Hugenschmidt2016ssr} in non-destructive ways. Complemented with polarized electron beams in the future electron-positron ($e^-e^+$) linear collider, high-energy polarized positron beams offer new prospects for stringently testing the Standard Model and efficiently suppressing unwanted background processes \cite{Moortgat2008pr,List2020arxiv}.


Low-energy and low-flux longitudinally polarized positrons can be produced from radioactive sources \cite{Zitzewitz1979prl}, but with large energy spreads and wide angular distributions. High-energy and initially unpolarized positrons can be directly polarized in the transverse direction in storage rings via radiative polarization \cite{Sokolov1968qed,Baier1972spu} (so-called Sokolov-Ternov effect), in which the positron spin is gradually aligned with the magnetic field direction. However, the buildup time of the polarization in most facilities is long, typically at a time scale of minutes to hours. In conventional accelerators, polarized positron beams are mainly achievable by the polarization transfer based on two steps. First a GeV-level unpolarized electron beam is passed through a helical undulator \cite{Alexander2008prl} or scattered off by a circularly polarized laser pulse \cite{Omori2006prl} to emit circularly polarized $\gamma$ photons, and subsequently some of these $\gamma$ photons further convert into longitudinally polarized $e^-e^+$ pairs in a high-Z target via Bethe-Heitler (BH) process. Another polarization transfer from electrons to positrons at MeV energies by polarized bremsstrahlung radiation is also demonstrated \cite{Abbott2016prl}, with an efficiency of about $10^{-4}$ from electron to positron.

Compared with BH process, polarized positrons could be more effectively generated via multiphoton Breit-Wheeler (BW) process \cite{Breit1934pr} in the strong-field quantum electrodynamics (QED) regime \cite{Berestetskii1982qed,Baier1998qed,Ritus1985jslr,piazza2012rmp}. Producing abundant positrons  or dense $e^-e^+$ plasmas by means of state-of-the-art ultraintense lasers \cite{Cartlidge2018science,Danson2019hplse} is extensively explored theoretically \cite{Bell2008prl,Sokolov2010prl,Nerush2011prl}. Unfortunately, these positrons are usually unpolarized due to the symmetrically oscillating field of common multi-cycle laser pulses \cite{Kotkin2003prab,Karlovets2011pra}. Recently, several schemes are proposed to construct asymmetric fields for the transversely polarized positron generation, e.g., one can employ elliptically polarized \cite{Wan2020plb} or two-color linearly polarized \cite{Chen2019prl} laser pulses. Longitudinally polarized positrons can be produced via the helicity transfer from initially longitudinally polarized electrons in ultraintense circularly polarized laser fields \cite{Li2020prl2}. Alternatively, the strong self-generated fields of high-energy, high-density electron beams in beam-beam collisions based on conventional accelerators can also trigger various QED processes \cite{Yokoya1992proc}, including multiphoton beamstrahlung \cite{Noble1987nimprsa,Chen1992prd,Tamburini2019arxiv}, multiphoton BW process \cite{Chen1989prl,Gaudio2019prab}, and even fully nonperturbative processes \cite{Yakimenko2019prl}. Whether one can obtain high-energy polarized positrons via multiphoton BW process from beam-beam collisions demands to study.

\begin{figure}[t]
\centering
\includegraphics[width=\mysize in]{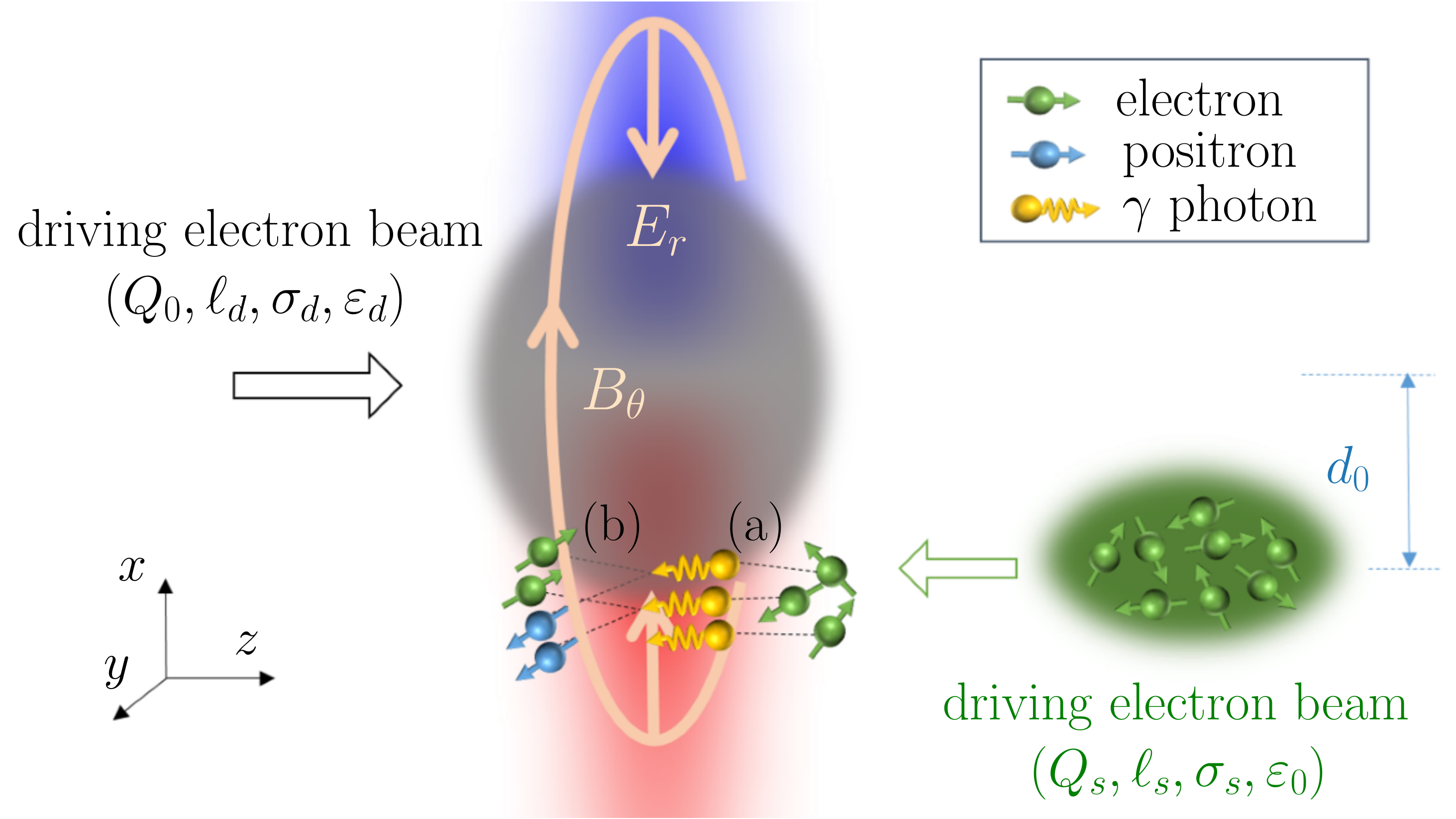}
\caption{\label{fig1} A schematic of generating polarized positron beams via the collision of two electron beams. An ultrarelativistic seeding beam collides head-on with a high-density driving beam. A transverse deviation distance of the beams results in strong unipolar field experienced by the seeding beam, whose magnetic field is along the $+y$ axis. There are two dominant QED processes in the interaction: (a) electrons and positrons emit $\gamma$ photons via multiphoton beamstrahlung; (b) $\gamma$ photons decay into polarized $e^-e^+$ pairs via multiphoton BW process.}
\end{figure}

In this article, we put forward a scheme for the generation of transversely polarized positron beams based on the collision of two electron beams. A high-density driving electron beam (total charge of several nC, beam size less than $1~\mu$m, which could be provided by FACET II \cite{Yakimenko2019prab}) provides the strong self-generated field, and the other ultrarelativistic seeding electron beam (kinetic energy of $\sim 10$ GeV) travels through part of the field for the $\gamma$ photon emission and the polarized $e^-e^+$ pair production, as sketched in Fig.~\ref{fig1}. The self-generated field of the driving beam, composed of the radial electric component and azimuthal magnetic component (of the order of $10^5$ T), is non-oscillating. By adjusting the transverse deviation distance between two beams properly, the seeding beam can only undergo a field which is unipolar. Hence, the magnetic field in the rest frame of the seeding beam is approximately along one direction, which is a key factor for the spin polarization of both primary electrons of the seeding beam and newly generated $e^-e^+$ pairs. With $e^-e^+$ spin and $\gamma$-photon polarization effects considered, our simulations present a highly polarized positron beam with polarization above $40\%$ or $60\%$ can be obtained if an initially unpolarized or specifically polarized seeding beam is employed.

This article is structured as follows. Section \ref{theory} presents the characteristics of the self-generated field of the driving beam and introduces the adopted theoretical model for the calculation of spin- and polarization-resolved pair production. In Sec.~\ref{results}, we analyze the simulation results and investigate the impacts of two beam parameters on the positron yield and positron polarization in detail. In addition, a method for further improving the positron polarization by employing an initially polarized seeding beam is proposed. Finally, a brief conclusion is drawn in Sec. \ref{conclusion}.

\section{THEORETICAL MODEL}
\label{theory}

Here, we consider a high-density driving electron beam propagating along $+z$ axis, with a bi-Gaussian charge density distribution
\begin{eqnarray}\label{eq1}
\rho=\rho_0e^{-\frac{r^2}{2\sigma_d^2}}e^{-\frac{(z-z_0-vt)^2}{2\ell_d^2}},
\end{eqnarray}
where $\rho_0=\frac{Q_0}{(2\pi)^{3/2}\sigma_d^2\ell_d}$ is the peak charge density, $Q_0$ is the total charge, $\ell_d$ is the bunch length, $\sigma_d$ is the transverse size, $v_d\approx c$ is the beam velocity, and $r=\sqrt{x^2+y^2}$. For an ultrarelativistic electron beam of Lorentz factor $\gamma_d=1/\sqrt{1-v^2/c^2}\gg 1$, the self-generated field around the beam is composed of the radial electric field $E_r$ and the azimuthal magnetic field $B_\theta$, which can be well approximated as \cite{Bassetti1983ieee}
\begin{eqnarray}\label{eq2}
E_r(r,z,t)~&&\approx B_{\theta}(r,z,t)\nonumber\\
&&\approx 4\pi e\rho_0\frac{\sigma_d^2}{r}\left(1-e^{-\frac{r^2}{2\sigma_d^2}}\right){}e^{-\frac{(z-z_0-vt)^2}{2\ell_d^2}}.
\end{eqnarray}
Longitudinal field components almost vanish, i.e., $B_z\approx 0$, $E_z\approx 0$. At $r=r_m\approx1.585\sigma_d$ and $z=z_0+vt$, $B_\theta$ has a maximum strength
\begin{eqnarray}\label{eq3}
B_\theta^{\rm max}\approx 1.09\times 10^4\times\frac{Q_0[{\rm nC}]}{\sigma_d[\mu{\rm m}]\times\ell_d[\mu{\rm m}]}[{\rm T}].
\end{eqnarray}
If the seeding electron beam with a kinetic energy $\varepsilon_0$ collides head-on with the driving electron beam, the maximum value of the quantum parameter $\chi_e=(e\hbar/m^3c^4)|F_{\mu\nu}p^{\nu}|$ during the interaction is
\begin{eqnarray}\label{eq4}
\chi_e^{\rm max}\approx0.0095\times \frac{Q_0[{\rm nC}]\times\varepsilon_0[{\rm GeV}]}{\sigma_d[\mu{\rm m}]\times\ell_d[\mu{\rm m}]},
\end{eqnarray}
where $F_{\mu\nu}$ is the field tensor and $p^{\nu}$ is the electron four-momentum.

In order to determine the spin vector of the generated positron ${\bm S}_+$ in multiphoton BW process, the spin- and polarization-resolved probabilities of photon emission $d^2W_{\gamma}/d\varepsilon_\gamma dt$ \cite{Baier1998qed,Li2019prl,Li2020prl} and pair production $d^2W_{\pm}/d\varepsilon_+dt$ \cite{Baier1998qed,Wan2020plb,Chen2019prl,Wan2020prr,Li2020prl} based on local constant field approximation (LCFA) are employed, with the latter written as 
\begin{eqnarray}\label{eq5}
\frac{d^2W_{\pm}}{d\varepsilon_{+} dt}&=&\frac{\alpha m^2c^4}{\sqrt{3}\pi \hbar \varepsilon_\gamma^2}\left\{\left[\frac{\varepsilon_{+}^2+\varepsilon_{-}^2}{\varepsilon_{+}\varepsilon_{-}}-\xi_3\right]K_{\frac{2}{3}}(y)+{\rm Int}K_{\frac{1}{3}}(y)\right.\nonumber\\ 
&&\left.-\xi_1\frac{\varepsilon_\gamma}{\varepsilon_-}K_{\frac{1}{3}}(y)(\bm S_+ \cdot \bm e_1)\right.\nonumber\\
&&\left.+\xi_2\left[\frac{\varepsilon_\gamma}{\varepsilon_+}{\rm Int}K_{\frac{1}{3}}(y)+\frac{\varepsilon_{+}^2-\varepsilon_{-}^2}{\varepsilon_{+}\varepsilon_{-}}K_{\frac{2}{3}}(y)\right](\bm S_+ \cdot \bm e_v)\right.\nonumber\\ 
&&\left.-(\frac{\varepsilon_\gamma}{\varepsilon_+}-\xi_3\frac{\varepsilon_\gamma}{\varepsilon_-})K_{\frac{1}{3}}(y)(\bm S_+ \cdot \bm e_2)
 \right\},
\end{eqnarray}
where $K_{\nu}(y)$ is the second-kind modified Bessel function of the order of $\nu$, ${\rm Int}K_{\frac{1}{3}}(y) \equiv \int_{y}^{\infty}K_{1/3}(x)dx$, $y=2\varepsilon_\gamma^2/(3\chi_\gamma\varepsilon_{+}\varepsilon_{-})$, $\varepsilon_-$ and $\varepsilon_+$ are the energies of the produced electron and positron, and $\varepsilon_\gamma$ is the parent $\gamma$ photon energy, respectively. Another quantum parameter $\chi_\gamma=(e\hbar^2/m_e^3c^4)|F_{\mu\nu}k^{\nu}|$ characterizes the pair production, where $\hbar k^{\nu}$ is the photon four-momentum. Variables $\xi_1$, $\xi_2$ and $\xi_3$ are Stokes parameters in the pair production frame ($\bm e_1, \bm e_2, \bm e_v$), where $\bm e_v$ is unit vector along photon velocity, $\bm e_1$ is unit vector along ${\bm E}+\bm e_v \times \bm B-\bm e_v \cdot (\bm e_v \cdot \bm E)$, and $\bm e_2=\bm e_v \times \bm e_1$. In general, the Stokes parameters $\xi_1$ and $\xi_3$ need to be calculated in real time through the matrix rotation if the external field experienced by $\gamma$ photons changes. In our cases, these two parameters nearly keep constant, since $\gamma$ photons are well collimated along $-z$ axis. Therefore, the directions of $\bm e_1$ and $\bm e_2$ are almost unchanged for each $\gamma$ photon during passing through the strong self-generated field of the driving beam. The small difference is ignored in our analysis for simplicity although it is captured in our simulations.

For the widely studied cases with laser fields, LCFA is generally considered to be valid under the strong-field condition of $a_0=|e|E_0/(mc\omega_0)\gg 1$ \cite{Ritus1985jslr,Piazza2018pra,Piazza2019pra,Ilderton2019pra,Podszus2019prd}, where $\omega_0$ is the laser frequency. The self-generated field of the driving electron beam can be approximately treated as the half-cycle laser field with a wavelength of $4\ell_d$. Therefore, the equivalent condition of LCFA is $a_0^*\approx2|e|E_r^{\rm max}\ell_d/\pi mc^2\gg 1$, i.e.,
\begin{eqnarray}\label{eq6}
a_0^*\approx4\times\frac{ Q_0[{\rm nC}]}{\sigma_d[\mu{\rm m}]}\gg 1.
\end{eqnarray}

\section{Simulation Results and analysis}
\label{results}

\subsection{Simulation setups}

To investigate the generation of polarized positrons in the collision of two ultrarelativistic electron beams, three-dimensional (3D) particle simulations are performed, in which we ignore the self-generated field of the low-density seeding beam (of the order of $10^2$ T, much weaker than that of the driving beam). The photon emission and pair production of quantum stochastic
nature are calculated using the standard Monte-Carlo algorithms \cite{Elkina2011prab,Ridgers2014jcp,Gonoskov2015pre} with spin- and polarization-resolved probabilities \cite{Baier1998qed,Li2019prl,Li2020prl,Wan2020plb,Chen2019prl,Li2020prl2}. The dynamics of electrons or positrons in the external electromagnetic field are described by classical Newton-Lorentz equations, and the spin precession is calculated according to the Thomas-Bargmann-Michel-Telegdi equation \cite{Bargmann1959prl}.

We take the driving electron beam with $Q_0=5\rm~nC$, $\ell_d=0.3~\mu$m, $\sigma_d=0.5~\mu$m, and initial kinetic energy $\varepsilon_d=1$ GeV. Note that $\varepsilon_d$ does not affect the self-generated field according to Eq.~(\ref{eq2}), if the precondition $\gamma_d\gg 1$ holds. Another initially unpolarized seeding electron beam propagating along $-z$ axis has the total charge $Q_s=20\rm~pC$ (consisting of $N_0\approx1.25\times 10^8$ primary electrons), bunch length $\ell_s=1.0~\mu$m, transverse size $\sigma_s=0.25~\mu$m, $\varepsilon_0=15$ GeV, and energy spread $\Delta\varepsilon_0/\varepsilon_0=0.01$, which can be provided by conventional accelerators or wakefield accelerations in the future \cite{Yakimenko2019prab,Pukhov2018prl}. There is a transverse deviation distance $d_0=0.8~\mu$m between two beams, i.e., $d_0\approx r_m$. At initial time $t=0$, the driving and seeding beams are centered at $z=-2~\mu$m and $2~\mu$m, respectively. For the chosen parameters, the maximum value of quantum parameter $\chi_e^{\rm max}$ during the interaction is 4.75, implying the interaction enters the QED regime, in which multiphoton BW process is important. Meanwhile, the normalized field strength $a_0^*\approx 40\gg 1$ according to Eq.~(\ref{eq6}) ensures the validity of LCFA.

\begin{figure}[t]
\centering
\includegraphics[width=\mysize in]{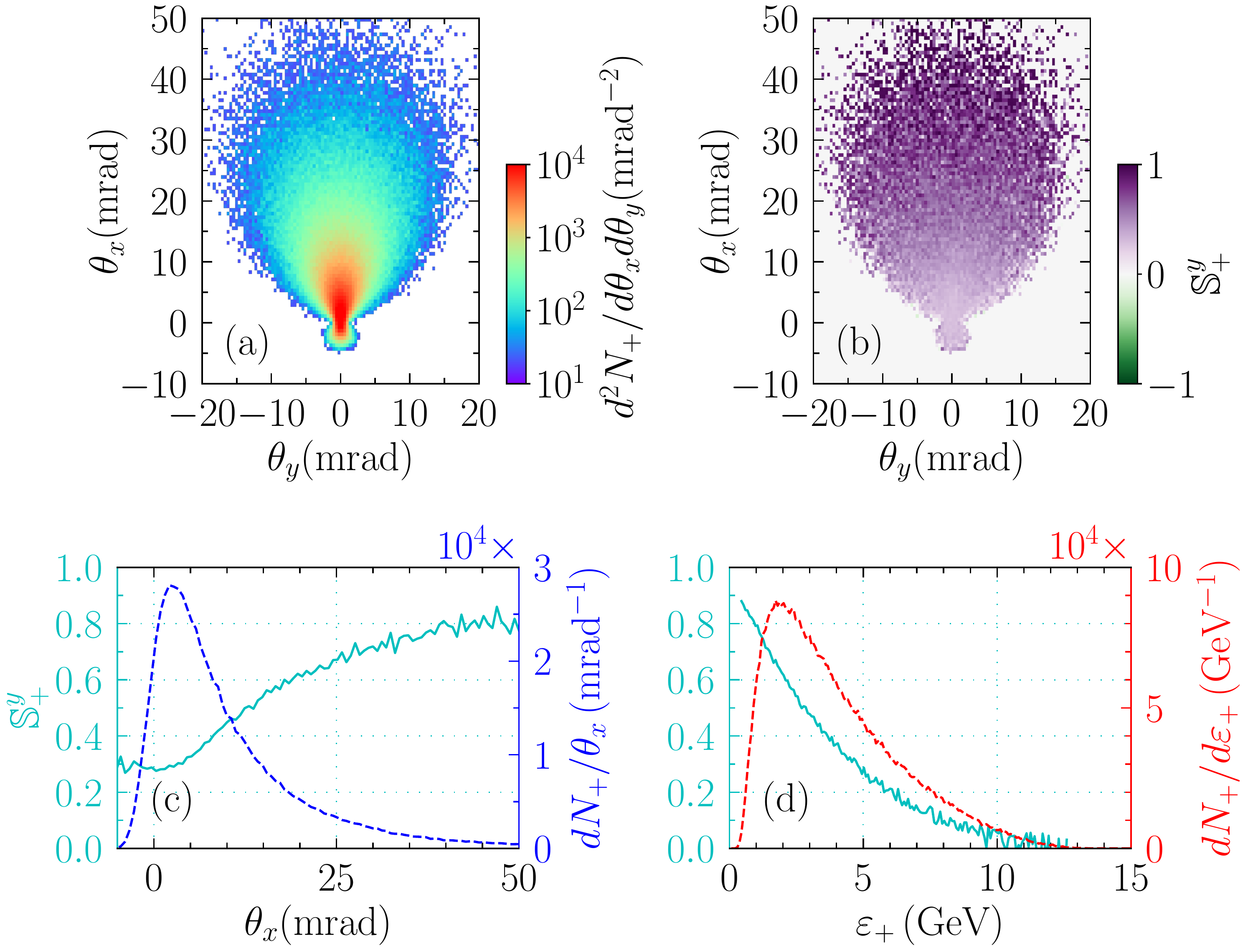}
\caption{\label{fig2} Properties of the generated positrons after the collision. Transverse angular distributions of (a) positron density $dN_+/d\theta_xd\theta_y$ and (b) positron polarization $\mathbb{S}_+^y$ vs deflection angles $\theta_x$ and $\theta_y$, where $\theta_{x(y)}=\arctan[p_{x(y)}/p_z]$. (c) $\mathbb{S}_+^y$ (cyan solid) and $dN_+/\theta_x$ (blue dashed) vs $\theta_x$. (d) $\mathbb{S}_+^y$ (cyan solid) and positron spectrum $dN_+/\varepsilon_+$ (red dashed) vs positron energy $\varepsilon_+$. }
\end{figure}

\subsection{Simulation results}

\begin{figure}[t]
\centering
\includegraphics[width=\mysize in]{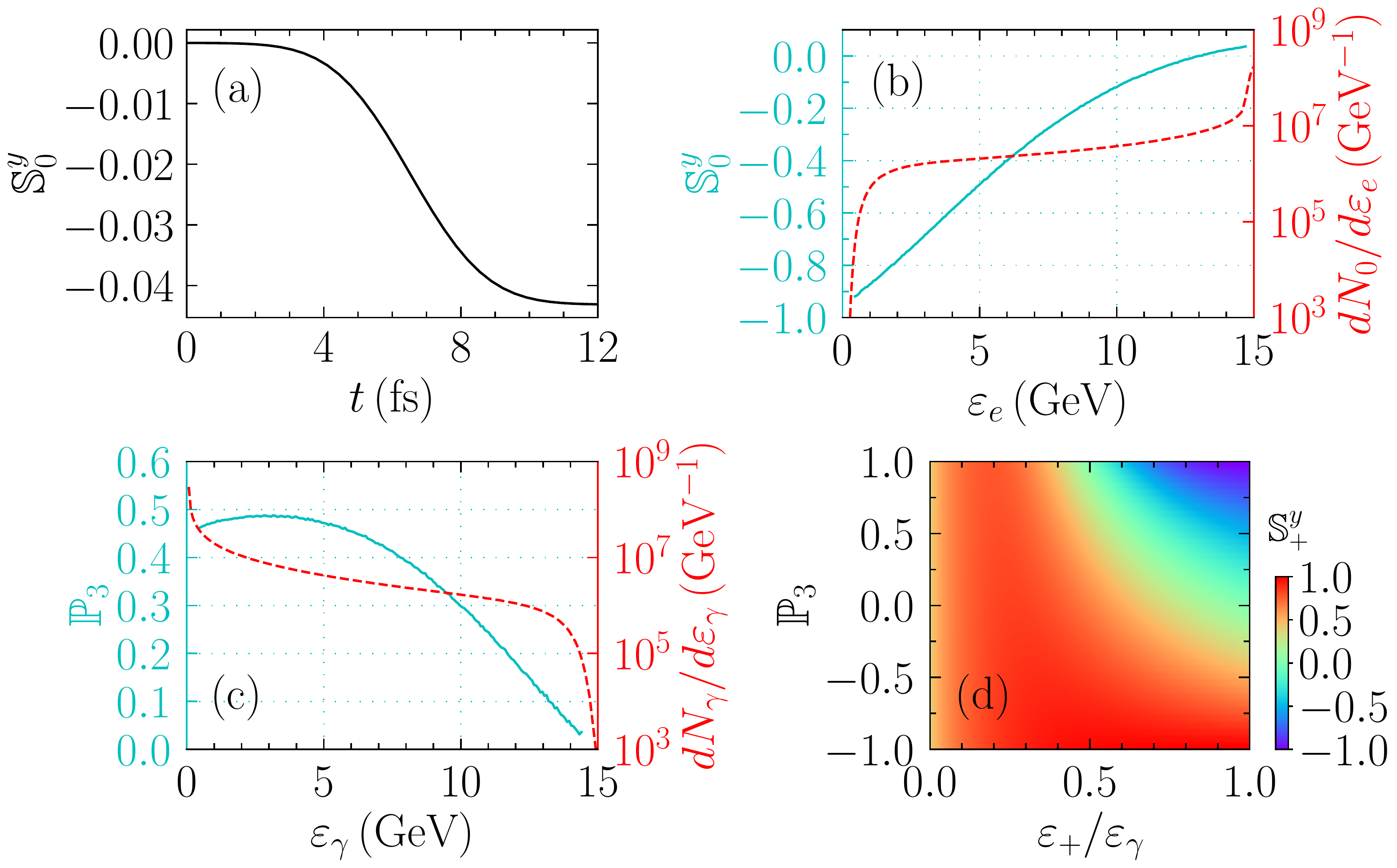}
\caption{\label{fig3} (a) Evolution of polarization degree of the seeding beam. (b) Electron spectrum $dN_0/d\varepsilon_e$ (red dashed) and electron polarization $\mathbb{S}_0^y$ (cyan solid) vs electron energy $\varepsilon_e$ of the seeding beam. (c) Photon spectrum $dN_\gamma/d\varepsilon_\gamma$ (red dashed) and photon polarization $\mathbb{P}_3$ (cyan solid) vs photon energy $\varepsilon_\gamma$. (d) Theoretical positron polarization $\mathbb{S}_+^y$ vs photon polarization $\mathbb{P}_3$ and the energy ratio $\varepsilon_+/\varepsilon_\gamma$ with $\chi_\gamma=1$. }
\end{figure}

We first present the essential properties of generated positrons in Fig.~\ref{fig2} and then analyze the spin- and polarization-dependent photon emission and pair production in Fig.~\ref{fig3}. Figure~\ref{fig2}(a) shows the transverse angular distribution of positron density $dN_+/d\theta_xd\theta_y$ after the collision. The generated positrons are mainly propagating along the initial seeding beam velocity ($-z$ axis) with an angular divergence around 10 mrad. The total positron number $N_+$ is about $4\times 10^5$, i.e., $N_+/N_0\approx 3.2\times 10^{-3}$. After generated, positrons are deflected along $+x$ axis by the strong field of the driving beam. Lower-energy positrons have larger deflected angles $\theta_x$ due to smaller relativistic masses. As a key result, the generated positrons appear to be transversely polarized, as shown by the corresponding positron polarization $\mathbb{S}_+^y$ (average positron spin $\overline S_+^y$) in Fig.~\ref{fig2}(b). The polarization degree of all generated positrons is about 0.43, and its positive value denotes that the positron spin is predominantly oriented along the $+y$ axis, which is the direction of the magnetic field in the rest frame of positrons. Positron polarization $\mathbb{S}_+^y$ increases with the increase of $\theta_x$, but at the same time the positron density dramatically declines, as shown in Fig.~\ref{fig2}(c). More specifically, at the larger angle $\theta_x>20$ mrad, $\mathbb{S}_+^y$ can exceed 0.6; at the smaller angle $\theta_x<10$ mrad where most positrons are concentrated, $\mathbb{S}_+^y$ is less than 0.4. In Fig.~\ref{fig2}(d), it clearly demonstrates that $\mathbb{S}_+^y$ is strongly dependent on the positron energy $\varepsilon_+$. For relatively low-energy positrons at 1 GeV, their $\mathbb{S}_+^y$ is about $0.8$, while $\mathbb{S}_+^y$ is only about $0.1$ for high-energy positrons at 10 GeV. The generated electrons have similar properties as those of positrons but with the opposite deflection angle (along $-x$ axis) and opposite polarization direction (along $-y$ axis) \cite{Baier1998qed,Wan2020plb,Chen2019prl}.

The initially unpolarized seeding beam can be directly polarized by the strong field of the driving beam via radiative polarization, as shown in Fig~\ref{fig3}(a). However, its final polarization degree is limited to $\mathbb{S}_0^y\approx -0.04$, much smaller than that of the generated positrons $\mathbb{S}_+^y\approx 0.43$. This is because the radiative polarization is a weaker spin-dependent process compared with the positron polarization via nonlinear BW process \cite{Chen2019prl}. Therefore, the former needs more polarization time, and the ultrashort interaction time $\sim\ell_d/c\approx 1$ fs significantly limits it. Figure~\ref{fig3}(b) illustrates the dependence of the electron polarization $\mathbb{S}_0^y$ on the electron energy $\varepsilon_0$ of the seeding beam. Lower-energy electrons possess higher polarization, since the electron spin is more likely to flip to the direction antiparallel to the magnetic field in its rest frame (i.e., $-y$ axis) only when a high-energy photon is emitted \cite{song2019pra} and the electron simultaneously loses a substantial fraction of its energy.


The emitted photons are also highly polarized, as shown in Fig.~\ref{fig3}(c). The positive value of photon polarization $\mathbb{P}_3$ (average Stokes parameter $\overline\xi_3$) indicates that they are mainly linearly polarized along the $x$ axis. Furthermore, $\mathbb{P}_3$ is closely related to the photon energy $\varepsilon_\gamma$, in which $\mathbb{P}_3$ of higher-energy photons is generally smaller than that of lower-energy photons. This is attributed to the spin- and polarization-dependent radiation probability \cite{Li2020prl,Song2021arxiv}, where $\mathbb{P}_3$ of high-energy photons emitted by unpolarized or just weakly polarized electrons is strongly suppressed, while $\mathbb{P}_3$ of low-energy photons is insensitive to electrons' spin and always positive \cite{Song2021arxiv}.

\begin{figure}[t]
\centering
\includegraphics[width=\mysize in]{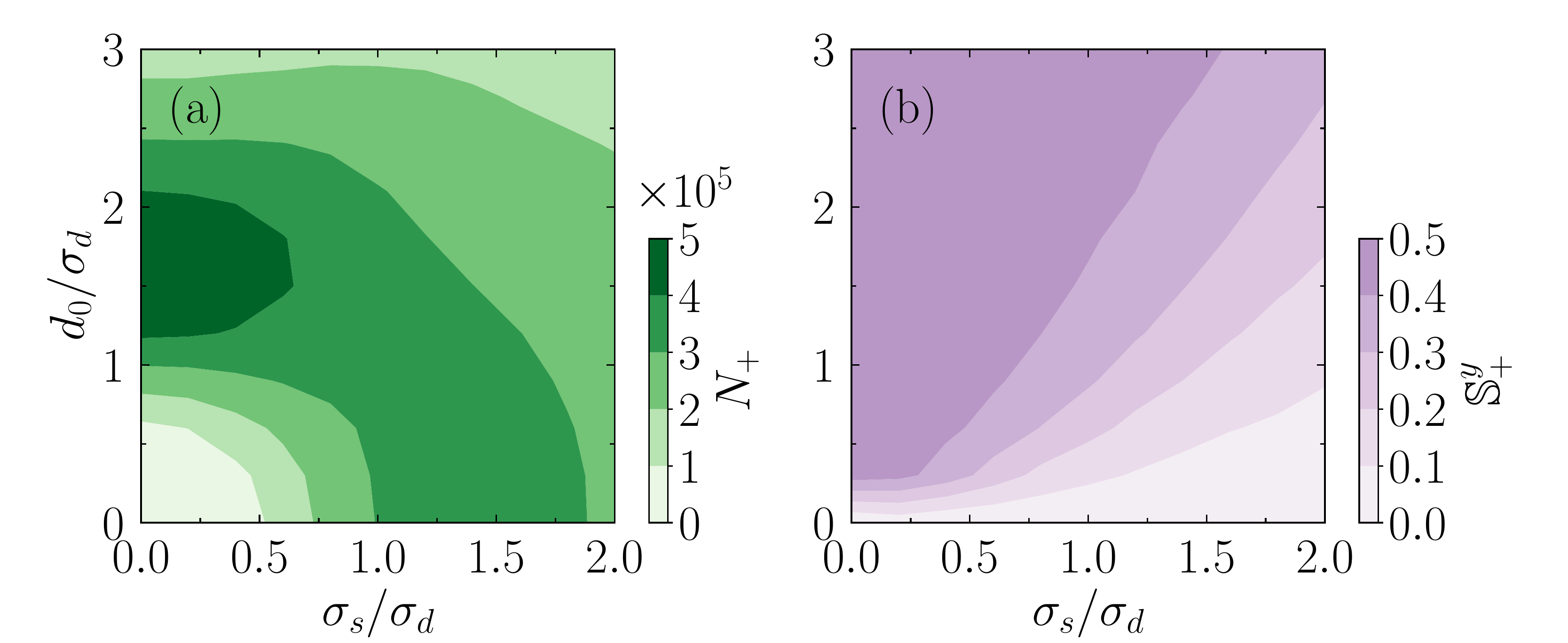}
\caption{\label{fig4} (a) Positron number $N_+$ and (b) positron polarization $\mathbb S_+^y$ vs $\sigma_s/\sigma_d$ and $d_0/\sigma_d$, where $\sigma_d$ is fixed at 0.5 $\mu$m and other parameter are the same as those in Fig.~\ref{fig2}.}
\end{figure}

Let us analyze the polarization process of generated positrons based on spin- and polarization-resolved pair production probability in Eq.~(\ref{eq5}). As discussed above, primary electrons of the seeding beam or generated electrons/positrons can only be transversely polarized, and consequently the emitted photons are mainly linearly polarized along the $x$ axis, i.e., $\overline\xi_1\approx 0$, $\overline\xi_2\approx 0$, and $\overline\xi_3\neq 0$. For this reason, the positron polarization $\mathbb{S}_+^y$ can be simplified as
\begin{eqnarray}\label{eq7}
\mathbb{S}_+^y=\frac{\varepsilon_\gamma(1/\varepsilon_+ -\mathbb{P}_3/\varepsilon_-)K_{\frac{1}{3}}(y)}{\left[\varepsilon_{+}/\varepsilon_{-}+\varepsilon_{-}/\varepsilon_{+}-\mathbb{P}_3\right]K_{\frac{2}{3}}(y)+{\rm Int}K_{\frac{1}{3}}(y)}.
\end{eqnarray}
According to Eq.~(\ref{eq7}), the theoretical $\mathbb{S}_+^y$ as a function of parent photon polarization $\mathbb{P}_3$ and the energy ratio $\varepsilon_+/\varepsilon_\gamma$ is plotted in Fig.~\ref{fig3}(d). One can notice that $\mathbb{S}_+^y$ of high-energy positrons is rather sensitive to $\mathbb{P}_3$, while it is insensitive for low-energy positrons. Moreover, since high-energy $\gamma$ photons are weak polarized, e.g., $\mathbb{P}_3<0.2$ for $\varepsilon_\gamma>12$ GeV as displayed in Fig.~\ref{fig3}(c), one can deduce that high-energy positrons have small $\mathbb{S}_+^y$ from Fig.~\ref{fig3}(d), which is consistent with the results shown in Fig.~\ref{fig2}(d). Noticeably, if intermediate $\gamma$ photons possess negative $\mathbb{P}_3$ (linearly polarized along $y$ axis), the polarization degree of generated positrons will be significantly improved, owing to positive $\mathbb{S}_+^y$ at both low and high energies. This can be achieved by employing an initially polarized seeding beam, which will be discussed below in more detail. 

\subsection{Impacts of two beam parameters}

We proceed to investigate impacts of two beam parameters on the positron number $N_+$ and polarization degree $\mathbb{S}_+^y$. To gain a highly polarized positron beam, the field experienced by the seeding beam should be unipolar. Hence, the seeding beam should keep away from the center of the driving beam. On the other hand, one needs to make sure that as many positrons as possible are generated. Therefore, there is an optimal transverse deviation distance $d_0$ between two beams for a given transverse size of the seeding beam $\sigma_s$. We present the results of achievable $N_+$ and $\mathbb{S}_+^y$ as a function of $d_0/\sigma_d$ and $\sigma_s/\sigma_d$ in Figs.~\ref{fig4}(a) and \ref{fig4}(b), respectively, where the transverse size of the driving beam $\sigma_d$ is fixed at 0.5 $\mu$m. Basically, the colliding condition of $\sigma_s=0.5\sigma_d$ and $d_0=1.5\sigma_d$ is optimal, which can simultaneously ensure appreciable $N_+$ and $\mathbb{S}_+^y$ in a more relaxed requirement for $\sigma_s$. Further decreasing $\sigma_s$ does not significantly increase $N_+$ or $\mathbb{S}_+^y$. In practice, if one cannot ensure this optimal condition, a larger $d_0$ should be adopted to meet a larger $\sigma_s$ when $\sigma_s/\sigma_d>0.5$. In that case, although $N_+$ decreases, $\mathbb{S}_+^y$ can be kept nearly unchanged around $0.4$.

\begin{figure}[t]
\centering
\includegraphics[width=\mysize in]{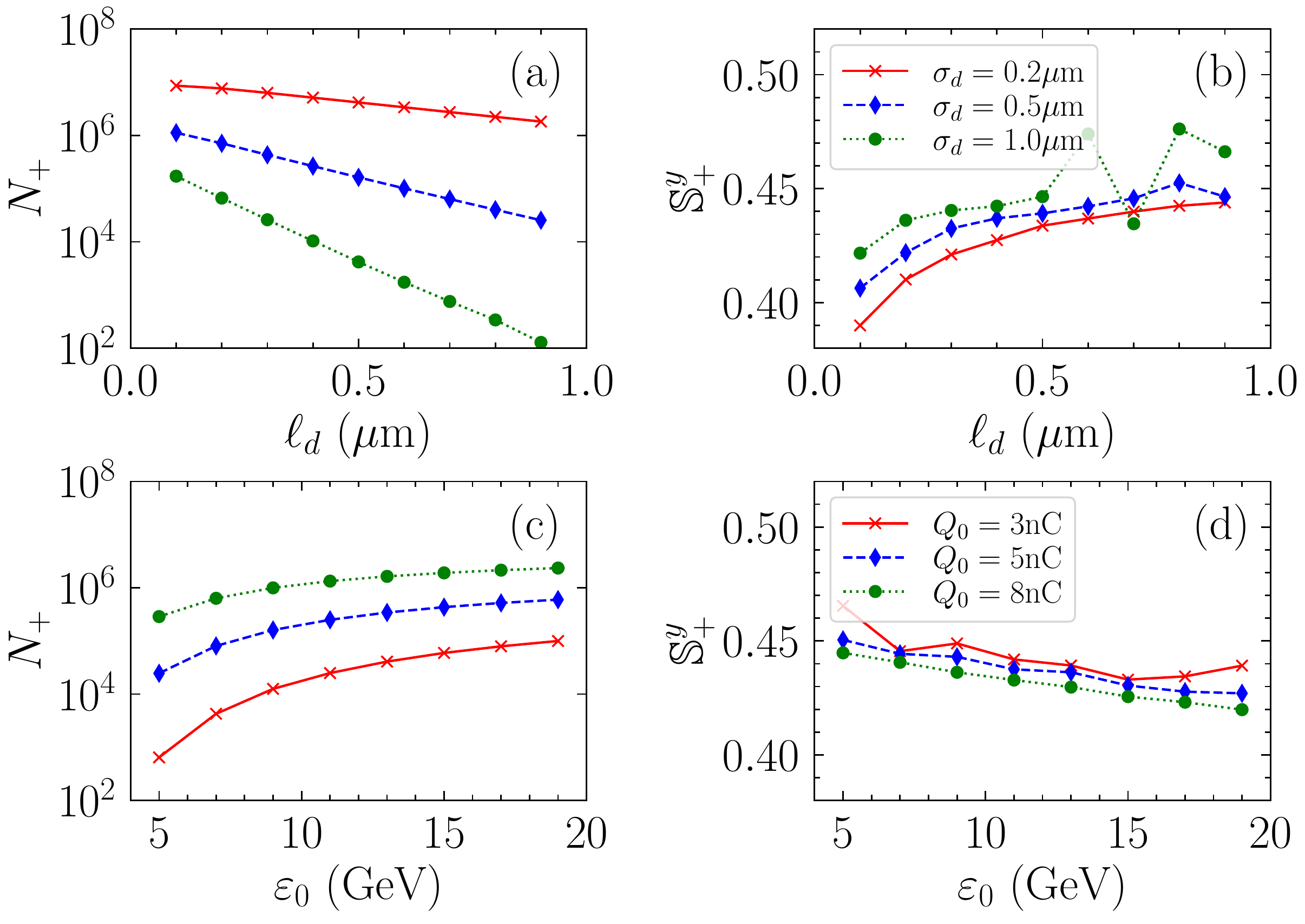}
\caption{\label{fig5} Top row: (a) positron number $N_+$ and (b) positron polarization $\mathbb{S}_+^y$ vs $\ell_d$ in the cases of $\sigma_d=0.2~\mu$m (red solid), $\sigma_d=0.5~\mu$m (blue dashed) and $\sigma_d=1.0~\mu$m (green dotted), respectively. Bottom row: (c) $N_+$ and (d) $\mathbb{S}_+^y$ vs $\varepsilon_0$ in the cases of $Q_0=3$ nC (red solid), $Q_0=5$ nC (blue dashed) and $Q_0=8$ nC (green dotted), respectively.}
\end{figure}

Then, in Figs.~\ref{fig5}(a) and \ref{fig5}(b), the roles of bunch length $\ell_d$ and transverse size $\sigma_d$ of the driving beam are investigated, under the optimal colliding condition obtained above and with a fixed total charge $Q_0=5$ nC. For each $\ell_d$, $N_+$ increases exponentially with the decrease of $\sigma_d$, while $\mathbb{S}_+^y$ decreases by about $10\%$ (from 0.45 to 0.4) within the range of scanning parameters. Hence, compressing driving beam is a very effective way to improve the yield of polarized positrons, due to the enhancement of the self-generated field, and consequently quantum parameter $\chi_e$ as well. For a highly compressed driving beam of $\ell_d=0.1~\mu$m and $\sigma_d=0.2~\mu$m, the generation efficiency can reach $N_+/N_0\sim0.1$. In addition, increasing the driving beam charge $Q_0$ or the seeding beam energy $\varepsilon_0$ can also dramatically increase $N_+$, as shown in Figs.~\ref{fig5}(c) and \ref{fig5}(d). $N_+$ is increased by 4 orders of magnitude from $\varepsilon_0=5$ GeV and $Q_0=3$ nC to $\varepsilon_0=20$ GeV and $Q_0=8$ nC, while the decrease of $\mathbb{S}_+^y$ is ignorable. For the former set of parameters, $\chi_e^{\rm max}$ is only 1, while for the latter, $\chi_e^{\rm max}$ is up to 7.6. In general, further compressing the driving beam size \cite{Jing2013prab,Yakimenko2012prl,White2018arxiv,Sampath2021prl} or increasing $Q_0$ or $\varepsilon_0$ can increase the positron yield effectively, meanwhile ensuring a high positron polarization, provided the optimal colliding condition of $\sigma_s=0.5\sigma_d$ and $d_0=1.5\sigma_d$ is fulfilled.

\subsection{Improving positron polarization by initially polarized seeding beam}

\begin{figure}[t]
\centering
\includegraphics[width=\mysize in]{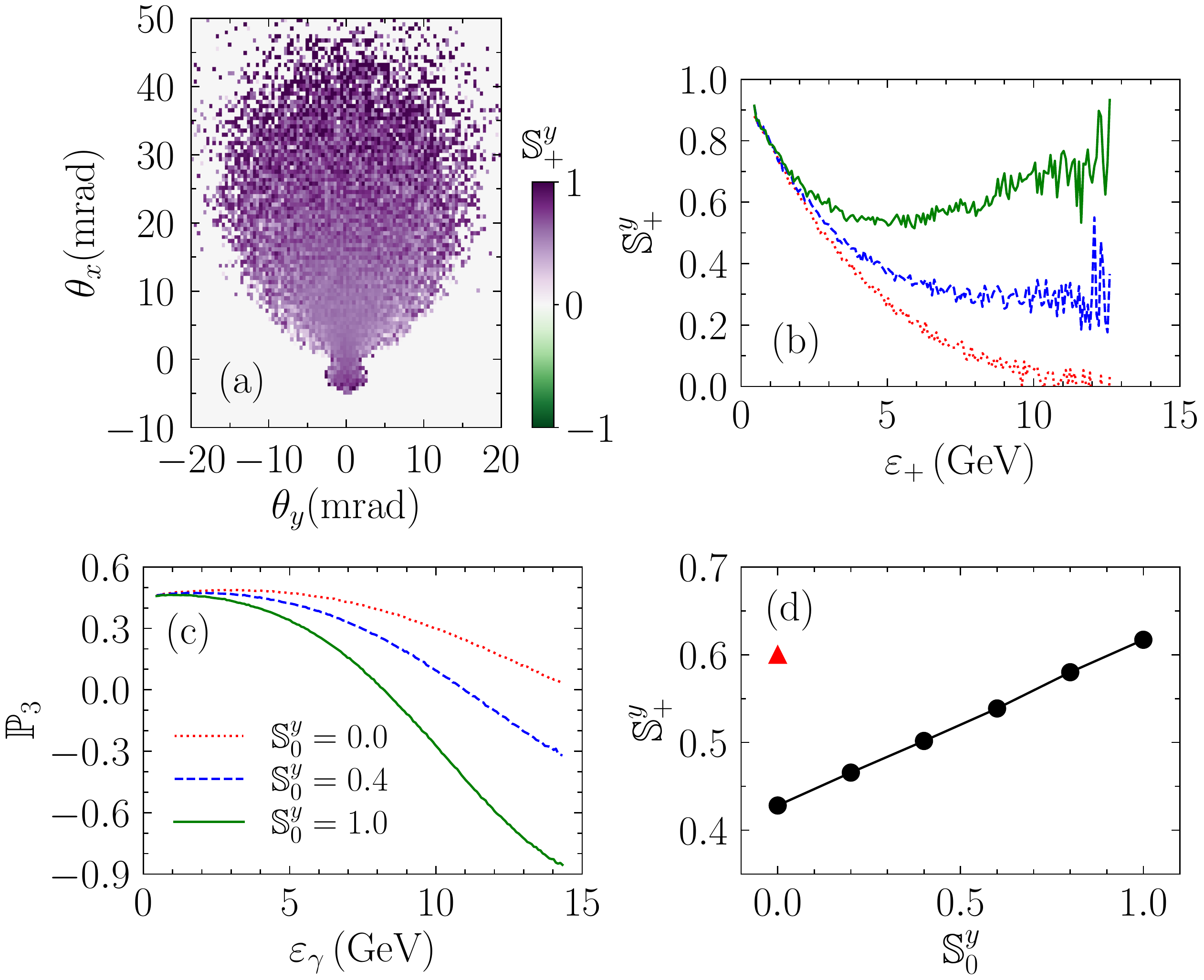}
\caption{\label{fig6} (a) Transverse angular distribution of positron polarization $\mathbb{S}_+^y$, in the case of an initially polarized seeding beam of $\mathbb{S}_0^y=1$. (b) $\mathbb{S}_+^y$ vs positron energy $\varepsilon_+$ and (c) photon polarization $\mathbb{P}_3$ vs photon energy $\varepsilon_\gamma$ under various initial polarization $\mathbb{S}_0^y=0$ (red dotted), $0.4$ (blue dashed), and $1$ (green solid). (d) Polarization degree of positrons vs $\mathbb{S}_0^y$. The red-triangle marker denotes the result excluding the effect of photon polarization. Other parameters are same as those in Fig.~\ref{fig2}.}
\end{figure}

As mentioned above, the polarization of high-energy positrons highly relies on the polarization of their parent $\gamma$ photons, while the latter depends on the initial polarization of the seeding beam [see Fig.~\ref{fig3}(d)]. For an unpolarized seeding beam ($\mathbb{S}_0^y=0$), the emitted $\gamma$ photons will carry positive $\mathbb{P}_3$, and $\mathbb{P}_3$ decreases with the increase of the photon energy $\varepsilon_\gamma$ [see Fig.~\ref{fig3}(c)]. To further improve the positron polarization, especially in the high-energy range, negative-$\mathbb{P}_3$ photons are required. Here, we propose to employ an initially polarized seeding beam with a positive polarization $\mathbb{S}_0^y>0$. We first take a $100\%$ polarized seeding beam of $\mathbb{S}_0^y=1$, whose spins directed parallel to the magnetic field in the rest frame of the seeding beam. The transverse angular distribution of the positron polarization $\mathbb{S}_+^y$ after the collision is shown in Fig.~\ref{fig6}(a). Compared with the unpolarized case in Fig.~\ref{fig2}(b), it is clearly visible that $\mathbb{S}_+^y$ has significantly improved in the small $\theta_x$ range, corresponding to high-energy positrons. In Fig.~\ref{fig6}(b), we plot $\mathbb{S}_+^y$ with respect to $\varepsilon_+$ under three different initial conditions of $\mathbb{S}_0^y=0$, $0.4$ and $1$, respectively. In the case of $\mathbb{S}_0^y=1$, one can see that $\mathbb{S}_+^y$ first decreases in the low-energy range of $\varepsilon_+<5$ GeV, and then increases with further increasing $\varepsilon_+$, which is quite distinct from continuously decreasing in the $\mathbb{S}_0^y=0$ case. Hence, one can obtain highly polarized positrons of high energies. The underlying reason is originated in the photon polarization. In Fig.~\ref{fig6}(c), we present the photon polarization $\mathbb{P}_3$ as a function of the photon energy $\varepsilon_\gamma$. For high-energy photons, which more probably decay into $e^-e^+$ pairs, their $\mathbb{P}_3$ is negative when $\mathbb{S}_0^y>1$. As predicted by Fig.~\ref{fig3}(d), these photons can generate high-energy polarized positrons of $\mathbb{S}_+^y>0$. Since $\mathbb{S}_+^y$ of low-energy positrons is also positive and insensitive to $\mathbb{P}_3$, the degree of positron polarization can be improved  substantially. The dependence of polarization degree of positrons on the initial polarization of the seeding beam is summarized in Fig.~\ref{fig6}(d). It can be seen that $\mathbb{S}_+^y$ monotonically increases with $\mathbb{S}_0^y$. In particular, $\mathbb{S}_+^y$ can exceed 0.6 for $\mathbb{S}_0^y=1$.

We also perform an additional simulation to further demonstrate the importance of the photon polarization. When the effect of the photon polarization is artificially switched off (or photon-polarization-averaged), the obtained $\mathbb{S}_+^y$ reaches 0.6 for the $\mathbb{S}_0^y=0$ case, as shown by the red-triangle marker in Fig.~\ref{fig6}(d), which is much larger than 0.43 of the case including the photon polarization effect. Hence, excluding the photon polarization will significantly overestimate the finial positron polarization, which is also pointed by \cite{Liu2020arxiv} that with the photon polarization effect included only about $30\%$ polarization of positrons can be achieved by the scheme utilizing two-color lasers \cite{Chen2019prl}.

\section{conclusion}
\label{conclusion}

In summary, through particle simulations, we have demonstrated that a high-energy transversely polarized positron beam with a polarization degree above $40\%$ can be generated via multiphoton BW process in the collision of two electron beams with a proper transverse deviation distance. Under the optimal colliding condition, the positron yield can be increased effectively while preserving a high polarization by increasing the charge density of driving beam (compressing beam size or increasing beam charge) or the energy of seeding beam. Utilizing the relationship of the polarization among primary electrons and generated $\gamma$ photons and positrons, the polarization degree of generated positrons can be improved to $60\%$ with an initially polarized seeding beam. Such highly transversely polarized positron beams could allow detecting possible sources of CP-violation \cite{Ananthanarayan2004prd} and testing specific triple-gauge couplings in the future $e^-e^+$ linear collider. Note that the positron polarization will be significantly overestimated provided the photon polarization effect is ignored.

\begin{acknowledgments}

This work was supported by the National Key R\&D Program of China (Grant No. 2018YFA0404801), National Natural Science Foundation of China (Grant Nos. 11775302 and 11827807), the Strategic Priority Research Program of Chinese Academy of Sciences (Grant Nos. XDA25050300, XDA25010300, and XDB16010200), and the Fundamental Research Funds for the Central Universities, the Research Funds of Renmin University of China (20XNLG01).
\end{acknowledgments}

\end{document}